\documentclass[prb,twocolumn,showpacs,superscriptaddress]{revtex4-1}
\usepackage{epsfig}
\usepackage{amsmath,amssymb}
\usepackage{graphicx}

\begin{document}

\title{Spin-orbit interaction and asymmetry effects on Kondo ridges at finite magnetic field}
\author{S. Grap,$^1$ S. Andergassen,$^1$ J. Paaske,$^2$ and V. Meden$^1$ \\
  {\small\em $^1$Institut f\"ur Theorie der Statistischen Physik, RWTH Aachen, 
    D-52056 Aachen, Germany} \\
  {\small\em and JARA-Fundamentals of Future Information Technology} \\
  {\small\em $^2$Nano-Science Center, Niels Bohr Institute, Universitetsparken 5, 2100 Copenhagen, Denmark}
}
\date{\small\today}
\begin{abstract}
We study electron transport through a serial double quantum dot with Rashba 
spin-orbit interaction (SOI) and Zeeman field of amplitude $B$ in presence of 
local Coulomb repulsion. 
The linear conductance as a function of a gate voltage $V_g$ equally shifting the levels 
on both dots shows two $B=0$ Kondo ridges which are robust against SOI as time-reversal 
symmetry is preserved. Resulting from the crossing of a spin-up and a spin-down level at 
vanishing SOI two additional Kondo plateaus appear at finite $B$. They are not protected 
by symmetry and rapidly vanish if the SOI is turned on. Left-right asymmetric 
level-lead couplings and detuned on-site energies lead to a simultaneous breaking 
of left-right and bonding-anti-bonding state symmetry. In this case the finite-$B$ 
Kondo ridges in the $V_g$-$B$ plane are bent with respect to the $V_g$-axis. 
For the Kondo ridge to develop different level renormalizations must be compensated by 
adjusting $B$.

\end{abstract}
\maketitle
\section{Introduction}

In linear response transport through quantum dots the spin Kondo effect shows up a 
as a plateau in the linear conductance $G$ 
when varying the level positions by an external gate voltage 
$V_g$, often referred to as a {\it Kondo ridge.}\cite{Hewson,Glazman,Ng,Goldhaber,Cronenwett,Schmid,Wiel} 
The width of the Kondo 
ridge is set by the local Coulomb interaction $U$ on the dot, which, in the Kondo 
regime, exceeds the 
level-lead hybridization $\Gamma$. The latter determines the width of the Lorentzian 
resonance in $G(V_g)$ at $U=0$. Breaking the two-fold Kramers-degeneracy 
by a local Zeeman field of amplitude $B$ destroys the Kondo ridge; along the $V_g$-axis 
the conductance plateau is split up into two Lorentzian resonances. 
In contrast, spin-orbit interaction (SOI), 
although breaking spin-rotational symmetry by designating a certain (spin) direction, 
does not destroy the Kondo effect.\cite{Meir,thesisB,jens} In the presence of SOI spin is no longer 
a good quantum number but a Kramers doublet remains as time-reversal symmetry is 
conserved. This leads to a Kondo effect in the presence of a local interaction provided
the gate voltage is tuned such that the dot is filled (on average) by an odd number of 
electrons (dominant spin fluctuations).  

In multi-level dots increasing $B$ 
might lead to energetically degenerate states 
(level crossings) resulting from different orbitals. If one is a spin-up state and 
one a spin-down state and the gate voltage is tuned such that an electron fluctuates 
between these states one might expect the emergence of a spin Kondo effect at {\it finite 
magnetic field}.\cite{Pustilnik,2,Izumida}
If the orbital quantum number 
is conserved in the leads in such systems additional orbital Kondo 
effects\cite{Cox} and combinations of spin and orbital Kondo effects\cite{Borda} may appear. 
Here we consider a setup where the orbital quantum number does not arise in the leads 
and we thus concentrate on spin Kondo effects.
In contrast to the standard $B=0$ Kondo effect the one appearing at finite $B$ is not protected 
by time-reversal symmetry and we show that it is {\it destroyed} in the presence of a 
finite SOI. 

\begin{figure}[t!]
\center{\includegraphics[width=65mm]{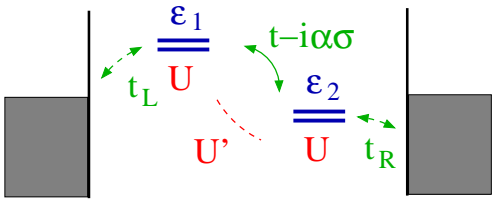}}
\caption{The considered setup consists of a serial double quantum dot with
  energies $\epsilon_{1/2}=V_g \pm \delta$ coupled by a hopping
  amplitude $t$ and a SOI of strength $\alpha$. The levels are split by an external Zeeman field $B$.
  The local Coulomb
  interaction is $U$ and the interaction between electrons on the two sites is $U'$.
  The system is coupled to noninteracting leads by hopping amplitudes $t_{L/R}$.}
\label{fig:fig1}
\end{figure}

We here study a serial double 
quantum dot, each having a single spin-degenerate level (at vanishing magnetic field) 
described by a tight-binding model with two lattice sites coupled by 
electron hopping of amplitude $t$ and connected to two semi-infinite 
noninteracting (Fermi liquid) leads via tunnel couplings
of strength $\Gamma_L$ and $\Gamma_R$. The on-site energies of the two levels 
are given by $\epsilon_{1/2}=V_g \pm \delta$.    
The Rashba SOI identifies the $z$-direction of the spin space and 
is modeled as an imaginary electron hopping with spin-dependent sign between the 
two lattice sites.\cite{Winkler,Mireles,BM,thesisB} 
We here exclusively consider the coupling of a magnetic field to the spin 
degree of freedom (Zeeman term)  and  neglect its effect on the orbital motion.
The relevant component of the Zeeman field perpendicular to the SOI defines
(without loss of generality) the $x$-direction.  
The local Coulomb repulsion is modeled as an on-site $U$ as well as a nearest-neighbor $U'$ repulsion 
and treated within an approximate static functional-renormalization group (fRG) 
approach.\cite{KEM} Our model is sketched in  Fig.~\ref{fig:fig1}.

In the absence of SOI a finite-$B$ Kondo effect will lead to a Kondo ridge. This conductance 
plateau can be detected if $G$ is computed (or measured) as a function of $V_g$ and $B$. We show 
that the finite-$B$ conductance plateau develops on a line 
parallel to the $V_g$-axis, the only exception being the case with broken 
{\it left-right} and {\it bonding-anti-bonding state symmetry} realized 
for $\Gamma_L \neq \Gamma_R$ {\it and} $\delta \neq 0$. 
In this case the finite-$B$ Kondo ridges are {\it bent} with respect to the $V_g$-axis; to follow 
the conductance maximum when changing $V_g$ one has to adjust $B$ to compensate
for the asymmetry-induced level renormalization.
As the two symmetries are generically broken in experimental systems our
results are important for the understanding of measurements of the
linear conductance of multi-level quantum dots as a function of $V_G$
and $B$.

This paper is organized as follows. In the next section, we introduce our double-dot 
model. In Sec.~\ref{sec:RG} details of the approximate fRG treatment of the Coulomb 
interaction specific to the present model are discussed. For later reference  
we also give a brief account of the appearance of the Kondo ridge for a single 
dot within our approximation scheme.  In Sec.~\ref{sec:res} we discuss our results. We 
first describe the effects of the SOI on the spectrum of the noninteracting 
isolated double dot. Next we present our results for the linear conductance $G(V_g,B)$ 
considering the entire parameter space. Our work is summarized in Sec.~\ref{concl}.

\section{Model}\label{sec:model}

Our multi-level quantum dot model is realized by a serial double dot 
each having a single spin-degenerate level (at $B=0$) as sketched in Fig.~\ref{fig:fig1}. 
The Hamiltonian of the isolated dot contains several terms
\begin{equation}
H_{\rm dot}=H_0+H_{\rm SOI}+H_Z+H_{\rm int} \;.
\end{equation}
The free part 
\begin{equation}
H_0=\sum_{\sigma} \left[\sum_{j=1,2} \epsilon_j d_{j,\sigma}^\dagger
  d_{j,\sigma}-t\left(
d_{2,\sigma}^\dagger d_{1,\sigma} +\mbox{H.c.}\right)\right] \; ,
\end{equation}
with $d_{j,\sigma}^\dagger$ being the creation operator of an electron
on the dot site $j=1,2$ (Wannier states) of spin $\sigma$, contains the 
conventional hopping $t>0$, and the on-site energies 
\begin{equation}
\epsilon_{1/2}=V_g \pm \delta
\end{equation} 
which can be tuned by an external gate voltage $V_g$. 
The difference of the on-site energies is parametrized 
by the level splitting $\delta$. The effect of SOI is taken into account by 
an imaginary hopping amplitude of spin-dependent sign. It is the 
lattice realization of a Rashba SOI resulting from spatial confinement.\cite{Winkler,Mireles,BM,thesisB}   
The Rashba hopping term with amplitude $\alpha>0$ reads
\begin{eqnarray}
H_{\rm SOI}&=&\alpha\sum_{\sigma,\sigma'}\left[ d_{2,\sigma}^\dagger
\left(i\sigma_z\right)_{\sigma,\sigma'} d_{1,\sigma'} + \mbox{H.c.} \right] \; ,
\end{eqnarray}
with the third Pauli matrix $\sigma_z$. Choosing the $z$-direction in spin 
space for the direction of the SOI breaks the spin-rotational invariance. 
The Zeeman field can have a component parallel to the SOI (that is in 
$z$-direction) and one perpendicular to it. For this we here choose (without 
loss of generality) the $x$-direction such that the (local) Zeeman term reads
\begin{eqnarray}
H_Z & = &   B \sum_{\sigma,\sigma'} \sum_{j=1,2} \left[  
d_{j,\sigma}^\dag (\sigma_z)_{\sigma,\sigma'} d_{j,\sigma'} {\rm sin} \, \theta  \right. \nonumber 
\\ && + \left. 
d_{j,\sigma}^\dag (\sigma_x)_{\sigma,\sigma'} d_{j,\sigma'} {\rm cos} \, \theta \right] \; .
\end{eqnarray}
For $\theta = \pm \pi/2$ the SOI and the $B$-field are (anti-)parallel. In this case 
the conventional hopping and the SOI can be combined to an effective hopping 
and the SOI does not have a specific effect. In particular, it does not destroy the 
finite-$B$ Kondo ridges (see Sec.~\ref{aa}). The local Coulomb interaction is included by
\begin{eqnarray}
H_{\rm int}&=&U\sum_{j=1,2} \left(n_{j,\uparrow} - \frac{1}{2} \right) \left( n_{j,\downarrow} - \frac{1}{2} \right) 
\nonumber \\ && +U' \left( n_{1} -1 \right) \left(n_{2}-1\right) 
\end{eqnarray}
for the local $U>0$ and nearest-neighbor $U'>0$ interactions respectively, with
$n_{j,\sigma}= d_{j,\sigma}^\dagger  d_{j,\sigma}$ and $n_j= \sum_\sigma n_{j,\sigma}$. 
By subtracting $1/2$ from 
$n_{j,\sigma}$ in the definition of $H_{\rm int}$ the point $V_g=0$ corresponds to half-filling
of the double dot even in the presence of Coulomb repulsion.

The dot Hamiltonian is supplemented by a term describing two semi-infinite 
noninteracting leads, which we here model as one-dimensional 
tight-binding chains 
\begin{eqnarray}
H_\mathrm{lead} =- \tau \sum_{\beta=L,R}  
\sum_{j=0}^\infty \sum_{\sigma} \left[ c_{\beta,j+1,\sigma}^\dagger 
c_{\beta,j,\sigma} +\mathrm{H.c.} \right]\;,
\end{eqnarray}
with lead operators $ c_{\beta,j+1,\sigma}^{(\dagger)}$ and equal band width 
$4 \tau$. In the following we choose $\tau$ as the unit of energy: $\tau=1$. 
The dot-lead couplings are given by a tunnel Hamiltonian
\begin{eqnarray}
\label{Hcoup}
H_\mathrm{coup}=&\sum\limits_{\sigma}& 
\big( t_L d_{1,\sigma}^\dagger c_{L,1,\sigma} + t_R c_{R,1,\sigma}^\dagger 
d_{2,\sigma} +\mathrm{H.c.} \big) \;,
\end{eqnarray}
with tunnel barriers set by $t_{L/R}$. We here consider the so-called wide band limit 
(see e.g. Ref.~\onlinecite{KEM}) in which the tunnel barriers only enter in combination 
with the local lead density of states (at lattice site 1) evaluated at the chemical potential 
by
$\Gamma_{L/R}= \pi t_{L/R}^2\rho_{\rm leads}$.

\section{Method: functional RG}\label{sec:RG}

\subsection{Flow equation for the self-energy}

We briefly review the applied approximation scheme which is based on the fRG method 
\cite{SalmhoferHonerkamp} focusing on aspects specific to the present model. 
A detailed description of the implementation for quantum-dot systems in 
the absence of SOI
is provided in Ref.~\onlinecite{KEM}. Recent extensions including SOI involve 
inhomogeneous quantum wires.\cite{BM}

Starting point of the fRG scheme is the bare ($U=U'=0$) Matsubara frequency 
propagator ${\mathcal{G}}_0$ 
of the double dot. The leads are projected onto the dot sites and enter via the 
hybridization $\Gamma_{L/R}$.\cite{KEM} In the basis 
\begin{eqnarray}
\label{basis}
\left\{ \left|1,\uparrow\right>, 
\left|1,\downarrow\right>, \left|2,\uparrow\right>, 
\left|2,\downarrow\right> \right\}
\end{eqnarray}
of single-particle states its inverse reads  
\begin{widetext}
 \begin{eqnarray}
&&\!\!\!\!\!\!\!\! {\mathcal{G}}_0^{-1}(i\omega) = \nonumber  \\ && \left( \begin{array}{cccc}
 i\omega-\epsilon_1-B \sin \theta + i \Gamma_L(\omega)\!\!\!\!& -B \cos \theta &t-i\alpha&0 \\
 - B \cos\theta&i\omega-\epsilon_1+ B \sin \theta + i \Gamma_L(\omega)\!\!\!\! &0&t+i\alpha \\
 t+i\alpha&0&i\omega-\epsilon_2-B \sin \theta+i \Gamma_R(\omega)\!\!\!\!&- B \cos \theta\\
0& t-i\alpha &-B{\rm cos} \, \theta& i\omega-\epsilon_2+B \sin \theta+i 
\Gamma_R(\omega)\end{array} \right) \label{g0}
 \; 
\end{eqnarray}
\end{widetext}
with $\Gamma_{L/R}(\omega)= \Gamma_{L/R} \, \mbox{sgn}(\omega)$. 
This propagator is replaced by one in which low-energy degrees of freedom below a cutoff $\Lambda$ are suppressed  
\begin{equation}
\mathcal{G}_0^\Lambda(i\omega)=\Theta(|\omega|-\Lambda)\mathcal{G}_0(i\omega)\;.
\end{equation}
The cutoff $\Lambda$ is later sent from $\infty$ down to $0$. 
Inserting $\mathcal{G}_0^\Lambda$ in the generating functional of the one-particle irreducible vertex 
functions, an infinite hierarchy of coupled differential equations is obtained by differentiating 
the generating functional with respect to $\Lambda$ and expanding it in powers of the external fields. 
Practical implementations require a truncation of the flow equation hierarchy.

We restrict the present analysis to the first order in the hierarchy and only consider the flow of the
one-particle vertex, that is the self-energy.  
It was previously discussed analytically\cite{AEM,KEM} that for a single dot the resulting flow equations 
capture the appearance of a Kondo ridge in $G(V_g)$ 
of width $\sim 1.5 \, U$.
The accuracy can be improved by including the flow of the static 
part of the two-particle vertex (effective interaction) which gives a Kondo plateau of width $\sim U$
in good agreement with the exact result.\cite{AEM,KEM}   
Here we are not interested in such quantitative improvements and instead keep the bare vertex. This 
has the advantage that the resulting flow equations for the matrix elements of the self-energy have a 
simple structure which not only allows for a fast numerical solution but provides the opportunity to 
gain analytical insights.  

The flow equation for the self-energy reads\cite{KEM} 
\begin{equation}
\frac{\partial}{\partial\Lambda}\Sigma_{a',a}^\Lambda=-\frac{1}{2\pi}\sum\limits_{\omega=\pm\Lambda}\sum\limits_{b,b'}e^{i\omega 0^+}{\mathcal{G}}_{b,b'}^\Lambda(i\omega)\Gamma_{a',b';a,b} \;, \label{DGLsigma}
\end{equation} 
where the indices $a,a',b,b'$ label the quantum numbers $({j,\sigma})$, $\Gamma_{a',b';a,b}$ is
the two-particle vertex, and 
the interacting Green function ${\mathcal{G}}$ is determined by the Dyson equation
\begin{equation}
\label{dyson}
{\mathcal{G}}^\Lambda(i\omega)=\left[\mathcal{G}_0^{-1}(i\omega)-\Sigma^\Lambda\right]^{-1} \; .
\end{equation}
The initial condition for $\Lambda_0 \to \infty$ is $\Sigma^{\Lambda_0} = 0$.\cite{KEM}
In the lowest-order scheme the two-particle vertex $\Gamma_{a',b';a,b}$ is given by the bare 
anti-symmetrized interaction. As the bare vertex is frequency independent, 
the approximate self-energy turns out to be frequency independent. 
Dynamical contributions are generated only at higher orders. As the latter are 
important for the conductance at temperatures $T>0$ the current approximation scheme 
is restricted to $T=0$. The correct temperature dependence of the (single-dot) Kondo 
ridge is only captured if the flow of a frequency dependent two-particle vertex---leading 
to a flowing frequency dependent self-energy---is kept.\cite{karrasch,Severin}
Within our approximation, the matrix elements of the self-energy 
at the end of the flow $\Sigma^{\Lambda=0}$ can be interpreted as 
interaction-induced  renormalizations to the noninteracting model parameters such 
as the SOI and conventional hopping amplitudes, as well as the on-site 
energies and the magnetic field.\cite{KEM}

\subsection{Computation of the linear conductance}

From the self-energy obtained at the end of the flow at $\Lambda=0$, the full propagator 
including interaction effects is determined via the Dyson equation 
(\ref{dyson}). From this various observables can be computed.\cite{KEM} 
Here we concentrate on the linear conductance. At $T=0$ current-vertex 
corrections vanish and the Kubo formula for the spin-resolved 
conductance assumes a generalized 
Landauer-B\"uttiker form\cite{Oguri} 
\begin{equation}
G_{\sigma,\sigma'} =\frac{e^2}{h} 
 |\mathcal{T}_{\sigma,\sigma'}(0)|^2 \; ,
\end{equation}
with the effective transmission $\mathcal{T}_{\sigma,\sigma'}(0)$ evaluated 
at the chemical potential.  For the present setup the transmission is given by the 
$(1,\sigma;2,\sigma')$ matrix element of the full propagator leading to\cite{KEM} 
\begin{equation}
G_{\sigma,\sigma'} = \frac{e^2}{h} 4 \Gamma_L \Gamma_R 
\left| \mathcal{G}_{1,\sigma;2,\sigma'}(0)\right|^2 \; .
\end{equation}
 
\subsection{Single-level quantum dot}

Before analyzing the serial double dot, for later reference we 
briefly discuss the single-level quantum dot within our 
approximation scheme.\cite{AEM,KEM}
In this case the flow equation for the effective (flowing) level 
position $V_\sigma^\Lambda = V_g + \sigma B + \Sigma_\sigma^\Lambda$ is
\begin{equation}
\label{eq:diffkondo}
  \frac{d}{d\Lambda} V_\sigma^\Lambda = \frac{UV_{\bar\sigma}^\Lambda/\pi}
  {(\Lambda+\Gamma)^2+(V_{\bar\sigma}^\Lambda)^2}\;,
\end{equation}
with initial condition $V_\sigma^{\Lambda=\infty} = V_g+\sigma B$, $\bar \sigma= - \sigma$,  
and $\Gamma=\Gamma_L+\Gamma_R$. 

At $B=0$ the level position is spin independent $V_\sigma^\Lambda = V^\Lambda$ 
and the differential equation can be solved 
analytically.\cite{AEM,KEM} For $U \gg \Gamma$ (in the Kondo 
regime) and $|V_g| \lesssim \,0.77 \, U$ the solution at $\Lambda=0$ is 
\begin{align}
 \label{eq:exp}
  V = V_g \exp \Big( -\frac{U}{\pi \Gamma}\Big) \; .
\end{align} 
It is this exponential pinning of the renormalized level position to 
zero (the chemical potential) which leads to the Kondo plateau in the total 
conductance $G(V_g,B)=G_\uparrow(V_g,B) +G_\downarrow(V_g,B) $ given by 
\begin{align}
  G(V_g,B=0) & = \frac{2e^2}{h} \, \frac{4 \Gamma_L \Gamma_R}{\Gamma^2}  
  \frac{\Gamma^2}{V^2 + \Gamma^2} \,.
\end{align}
For $U=0$ the level position $V=V_g$ is unrenormalized and $G$ reaches 
its maximum value $G_{\rm max} = (2e^2/h) \, 4\Gamma_L
\Gamma_R/\Gamma^2$ only at the resonance voltage $V_g=0$ (Lorentzian resonance of width $\Gamma$). 
For $U>0$, $V$ is pinned to zero around $V_g=0$ for a 
gate-voltage range of width $\sim 1.5 \, U$, where the Kondo plateau develops in the conductance  
$G\simeq G_{\rm max}$.    

We next consider a finite magnetic field $B \neq 0$. The flow equation for the effective  
$B^\Lambda =  ({V}^\Lambda_{\uparrow}-{V}_{\downarrow}^\Lambda )/2$
reduces to 
\begin{align}
   \frac{d}{d\Lambda} B^\Lambda = -\frac{U B^\Lambda/\pi}
  {(\Lambda + \Gamma)^2+(B^\Lambda)^2}\label{bren}
\end{align}
at $V_g=0$.
The renormalized magnetic field $B_{\rm ren}$ at $\Lambda=0$ hence shows the same exponential 
behavior (with prefactor $B$ instead of $V_g$) as the renormalized level 
position Eq.\ (\ref{eq:exp}) except for the \textit{reversed} 
sign in the exponent. This leads to a dramatic increase of the renormalized field $B_{\rm ren}$
compared to the bare one. 
For $U>0$ the total conductance at $V_g=0$ 
\begin{align}
  G(V_g=0,B) & = \frac{2e^2}{h} \, \frac{4 \Gamma_L \Gamma_R}{\Gamma^2}  
  \frac{\Gamma^2}{B_{\rm ren}^2 + \Gamma^2} \; ,
\end{align} 
in the $B$ direction becomes exponentially 
(set by the Kondo scale\cite{Hewson,AEM,KEM}) sharp, instead of Lorentzian-like of width $\Gamma$ for $U=0$. 
In fact, this holds not only at $V_g=0$ but for all gate voltages within the $B=0$ conductance plateau. 
Qualitatively the Kondo ridge of a single-level dot in the $V_g$-$B$ plane looks similar to 
the surrounding of one of the $B=0$ Kondo ridges appearing in our double dot model as e.g. shown 
in Fig.\ \ref{fig:fig2}.

\vspace{0.6cm}

\section{Results}\label{sec:res}

\subsection{The noninteracting isolated double dot}

For a detailed understanding of the Kondo ridges it is instructive to first 
study the noninteracting isolated double dot. 
In the basis of Eq.\ (\ref{basis}) the single-particle Hamiltonian $h_{\rm dot}$ is represented 
as a complex $4 \times 4$ matrix 
\begin{widetext}
 \begin{equation}
h_{\rm dot}=\left( \begin{array}{cccc}
 V_g+\delta+ B \sin \theta& B \cos \theta &-t+i\alpha&0 \\
 B \cos \theta&V_g+\delta-B \sin \theta &0&-t-i\alpha \\
 -t-i\alpha&0&V_g-\delta+B \sin\theta&B \cos \theta\\
0& -t+i\alpha & B \cos \theta& V_g-\delta-B \sin \theta\end{array} \right) \;.
\label{0}
\end{equation}
\end{widetext}
The corresponding eigenvalues are 
\begin{eqnarray}
\lambda& =&V_g\pm\sqrt{B^2+t^2+\delta^2+\alpha^2\pm2B\sqrt{t^2+\delta^2+\alpha^2{\rm sin}^2\theta}}\nonumber\\
&=&V_g\pm\sqrt{\left(B\pm\sqrt{t^2+\delta^2+\alpha^2{\rm sin}^2\theta}\right)^2+\alpha^2{\rm cos}^2\theta}\;.\nonumber\\
\label{e}
\end{eqnarray}
The spectrum is invariant under the transformation $B \to -B$ and symmetric at $V_g=0$.  
A finite on-site energy $\delta>0$ yields an effective hopping 
$t_{\rm eff}=\sqrt{t^2+\delta^2}$. For vanishing SOI ($\alpha=0$) Eq.~(\ref{e}) 
reduces to 
\begin{equation}
\label{EVform}
\lambda=V_g\pm\left(t_{\rm eff} \pm B\right) \; .
\end{equation}
Naturally, the 
$\theta$-dependence drops out. For $\theta=\pm \pi/2$ the SOI and the Zeeman field
are (anti-)parallel and $\alpha$ can be absorbed into an 
effective  hopping  $t_{\rm eff}=\sqrt{t^2+\delta^2 + \alpha^2}$. The eigenvalues
are then of the $\alpha=0$-form Eq.\ (\ref{EVform}).  
 
For the appearance of a spin Kondo effect (after turning on $\Gamma_{L/R}$ as well as $U$ and $U'$) 
it is necessary that two degenerate levels of opposite spin are located at zero energy 
(the chemical potential).  Zero energy spin-degenerate levels are obtained at $B=0$ and 
$V_g=\pm t_{\rm eff}$ (bonding and anti-bonding states). 
This will lead to the standard spin Kondo effect related to the presence of a (spin)
Kramers doublet when $U$ and $U'$ are switched on.
Increasing $B$ the spin-up level 
of the bonding state and the spin-down level of the anti-bonding one approach each other. 
For either $\alpha=0$ or $\theta=\pm \pi/2$ they become degenerate (cross) at zero 
energy for $B_c=\pm t_{\rm eff}$ and $V_g=0$. Besides the two $B=0$ Kondo ridges developing 
for all dot parameters, for $\alpha=0$ or $\theta=\pi/2$ one might thus expect the appearance of 
two finite-$B$ Kondo ridges. 
Viewed on the basis of the many-body energies of the isolated {\it interacting} double dot this situation corresponds to the crossing of a two-particle $S=1$, $S_z=-1$ state with a corresponding $S=1$, $S_z=0$ state. The resulting Kondo effect can thus also be referred to as a singlet-triplet one.
For $\alpha>0$ and an arbitrary angle $\theta$ between 
the SOI and the Zeeman field the finite-$B$ level crossings turn into avoided crossings and 
no zero-energy degeneracies are possible at finite $B$.

\subsection{Numerical results for the conductance}

We next present results for the linear conductance $G(V_g,B)$ obtained by numerically 
solving the flow equation (\ref{DGLsigma}). For our purposes it is sufficient 
to consider only a single set of $U,U'$ and $t$. The results depend only 
quantitatively on the strength of the local Coulomb interaction 
and the inter-dot hopping $t$, as long as  $U/\Gamma_{L/R}$ and $t/t_{L/R}$ are 
sufficiently large. We here focus on $U=U'=1$ and $t=1$. 

\subsubsection{Vanishing SOI}
\label{aa}

We start our discussion with the case of vanishing SOI. This is trivially
realized for $\alpha=0$. As discussed above, for 
$\theta=\pi/2$ the SOI can be absorbed into an effective hopping.
In this case the Zeeman field and the SOI are (anti-)parallel
and only a single Pauli matrix enters the Hamiltonian. The physics is thus similar 
to the $\alpha=0$ case.   

\begin{figure}[t!]
\center{\includegraphics[clip=true,width=8.75cm]{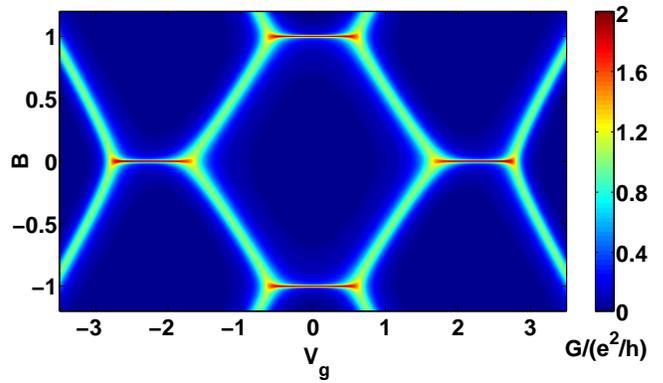}}
\caption{\label{fig:fig2} (Color online)
 Conductance $G(V_g,B)$ for $t=1$, $U=U'=1$, $t_{L/R}=0.3$, and $\delta=0$ in the 
absence of SOI ($\alpha=0$).}
\end{figure}

\begin{figure}[t!]
\center{\includegraphics[clip=true,width=9.25cm]{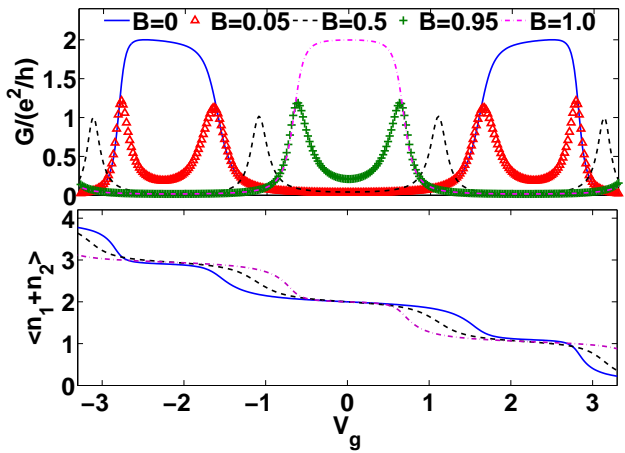}}
\caption{\label{fig:fig3} (Color online)
 Gate-voltage dependence of the conductance $G$ and dot occupation $\left< n_1+n_2 
\right>$ for different values of $B$ and the same parameters as in Fig.~\ref{fig:fig2}.}
\end{figure}

In Fig.~\ref{fig:fig2} we show $G(V_g,B)$ for a left-right symmetric 
setup with $t_{L/R}=0.3$ and $\delta=0$ in the absence of SOI ($\alpha=0$). 
The conductance shows the expected two pairs of $B=0$ and $V_g=0$ Kondo ridges. 
Due to the renormalization of the inter-dot hopping the 
$B=0$ plateaus are not centered around $V_g = \pm t_{\rm eff}=t=1$ but 
renormalized gate voltages. For the present parameters the renormalization of 
the inter-dot hopping and the magnetic field (almost) cancel each other such that the 
finite $B$ ridges are located at $B_c \approx  \pm t_{\rm eff}=t=1$. 
For $U' < U$, $B_c$ is renormalized to smaller values.
All Kondo plateaus exhibit the same maximal height of $2 e^2/h$. 
This can be understood from transforming the dot-lead 
coupling Eq.~(\ref{Hcoup}) into the basis of bonding ($b$) and anti-bonding ($\bar b$)
states of the noninteracting isolated dot.
The couplings between the leads and the two states are 
\begin{equation}
\Gamma_{b,L/R}=\frac{\sqrt{t^2+\delta^2}\mp \delta}{2\sqrt{t^2+\delta^2}}\,\Gamma_{L/R}\;,
\label{leadcoup}
\end{equation}
and $\Gamma_{\bar b,L/R}=\Gamma_{b,R/L} \,t^2_{L/R} / t^2_{R/L}$.
For the considered left-right symmetric case and $\delta=0$ 
the bonding and anti-bonding states have the same total coupling 
$\Gamma_{b/\bar b}=\Gamma_{b/\bar b,L}+\Gamma_{b/\bar b,R}$  
and the couplings are left-right symmetric implying unitary 
conductance. Increasing $|B|$ from $B=0$ the Kondo ridges are suppressed on the
exponential Kondo scale and resonance peaks of height $e^2/h$ and 
width $\Gamma$ develop. The position of the resonance peaks varies linearly 
with $V_g$. Eventually the peaks corresponding to the spin-up level of the bonding 
state and the spin-down level of the anti-bonding one merge with the 
finite-$B$ Kondo ridges. Figure \ref{fig:fig3} shows $G(V_g)$ at fixed 
different $B>0$ to further exemplify this. The shoulders appearing in the 
$B=0$ bonding (anti-bonding) state Kondo ridges are linked to the  
presence of the anti-bonding (bonding) one. Increasing the Coulomb interaction has the two obvious 
effects of broadening the Kondo ridges and increasing the distance between the 
centers of the plateaus; the latter also holds for increasing $t$. 
In the lower panel of Fig. \ref{fig:fig3} the respective dot fillings are shown
for various values of $B$.
For $B=0$ the dot occupation exhibits plateaus at odd fillings. Their width corresponds
to the Kondo ridges observed in the conductance. Similarly, for a finite magnetic field $B=1$ the plateau around $V_g=0$ in the conductance is reflected in the filling.

We note that $G(V_g,B)$ is symmetric 
with respect to $B \to -B$ (see the eigenvalues Eq.\ (\ref{e})) and
$V_g \to -V_g$. While the former symmetry is given by the Hamiltonian and 
holds for {\it all} parameter sets, the latter is specific to the parameters 
chosen here ($t_L=t_R$ and $\delta=0$). 
The conductance remains symmetric under $V_g \to -V_g$  if at least either $\delta=0$ 
or $t_L = t_R$ holds. In these cases the four Kondo ridges do no longer reach
the unitary value $2 e^2/h$ but exhibit an equally reduced conductance plateau as the bonding 
and anti-bonding states are coupled with the same asymmetry and the same total coupling 
to the leads.

\begin{figure}[t!]
\center{\includegraphics[clip=true,width=8.75cm]{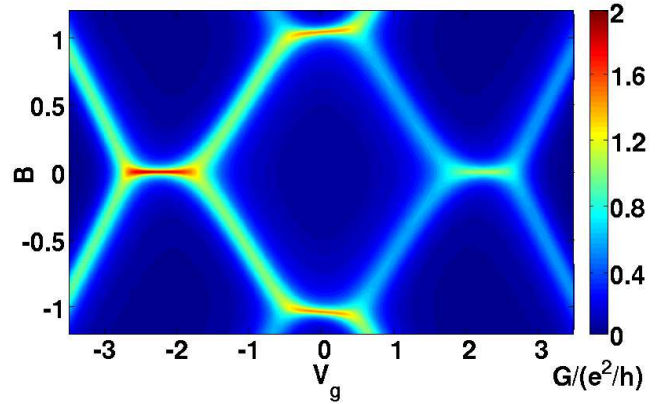}}
\caption{\label{fig:fig4} (Color online)
 Conductance $G(V_g,B)$ for $t=1$, $U=U'=1$, $\alpha=0$ 
 with asymmetric couplings to the leads  $t_L=0.3$, $t_R=0.5$ and finite level 
 splitting $\delta=-0.3$.}
\end{figure}

The most interesting situation arises if the left-right symmetry and the bonding-anti-bonding 
state symmetries are simultaneously broken. This is achieved for $t_L \neq t_R$ {\it and} 
$\delta \neq 0$. In this case the $V_g \to -V_g$ symmetry of $G(V_g,B)$ is broken as shown 
in Fig.~\ref{fig:fig4}. For our parameters the couplings of the anti-bonding state have a 
strong left-right asymmetry leading to a $B=0$  
Kondo plateau with significantly reduced conductance (around $V_g =2.2$). 
The bonding state has a weaker asymmetry such that the $B=0$ Kondo ridge 
centered around $V_g = -2.2$ almost reaches the 
unitary conductance. The total coupling $\Gamma_{\bar b}$ is larger than $\Gamma_{b}$. 
With the breaking of the $V_g \to -V_g$ symmetry, manifest already from the comparison 
of the two $B=0$ Kondo ridges, the finite-$B$ Kondo ridges (centered 
around $V_g=0$) are no longer necessarily parallel to the $V_g$-axis. 
In fact they are {\it bent} with respect to this axis as becomes apparent 
from Fig.~\ref{fig:fig4}. It turns out that the direction of bending is always 
away from the state with stronger total level-lead coupling. For our model this 
is the state with larger left-right asymmetry.  
This bending cannot be predicted considering the isolated double-dot even by
including the Coulomb interaction as it results from a level renormalization 
associated with the dot-lead couplings.  
Related level renormalizations are discussed 
in Refs.\ \onlinecite{Holm} and \onlinecite{Hauptmann}.
In experiments on multi-level quantum 
dots left-right symmetry is difficult to realize and the states at different 
energies will have different level-lead couplings. Therefore finite-$B$ spin 
Kondo ridges appearing in measurements are expected to be generically bent with
 respect to the $B=0$ ones. This result is relevant for the understanding 
of finite-$B$ Kondo ridges observed in conductance measurements on multi-level 
carbon nanotube quantum dots.\cite{kasper}

\subsubsection{Effect of the SOI}

\begin{figure}[t!]
\center{\includegraphics[clip=true,width=8.75cm]{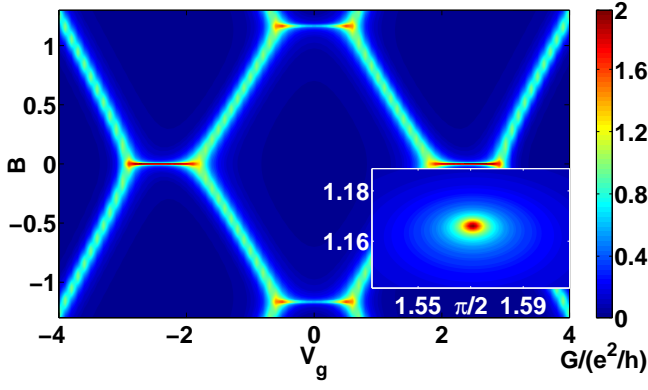}}
\caption{\label{fig:fig5} (Color online)
 Conductance $G(V_g,B)$ for $t=1$, $U=U'=1$, $t_{L/R}=0.3$, and  
$\delta=0$ with $\theta=1.56$ and finite SOI 
$\alpha=0.6$. Inset: Conductance $G(\theta,B)$ at $V_g$=0.}
\end{figure}

For the discussion of the effect of the SOI on the Kondo ridges we focus on the symmetric case 
with $t_L = t_R$ and $\delta=0$. Figure~\ref{fig:fig5} shows $G(V_g,B)$ 
for $\alpha = 0.6$ and an angle $\theta = 1.56$ close to the parallel configuration
with $\theta = \pi/2$. Thus the effective SOI given by the component perpendicular to 
the direction of the Zeeman field is small. For the noninteracting single-particle levels of 
the isolated double dot this implies a small minimal distance between the levels avoiding the 
crossing at finite $B$. Therefore remnants of the finite-$B$ Kondo ridges are still observable. The inset 
shows $G$ at $V_g=0$ as a function of $\theta$ and $B$. Increasing the SOI component
perpendicular to the direction of the Zeeman field by deviating from $\theta=\pi/2$ obviously
destroys the finite-$B$ Kondo effect as expected. This provides a way to probe the presence of SOI 
in multi-level dots: after observing finite-$B$ spin Kondo ridges one can probe their robustness 
by changing the direction of the magnetic field.    
Remarkably, the SOI introduces an intriguing angular dependence even 
for a magnetic field coupling exclusively to the spin degree of freedom.\cite{jens}
From the angular dependence of the gap opening in the single-particle 
energy spectrum the strength of the SOI can be extracted by spectroscopic measurements.

\subsection{Analytical insights}

We next provide analytical insights to our findings in the absence of SOI 
by analyzing the fRG flow equation (\ref{DGLsigma}). 
To simplify the analysis we focus 
on the case of purely local Coulomb interactions with $U'=0$. Due to the absence of
Fock terms $t$ remains unrenormalized, but this only yields 
quantitative changes compared to the results shown in the last subsection.

In the absence of SOI $\theta$ does not play any role and can be chosen arbitrarily. For $\theta=\pm \pi/2$ 
the matrix (\ref{g0}) is block diagonal and we can restirct the analysis to a single spin sector.
Introducing $V_{j\sigma}^{\Lambda}=\epsilon_j+\sigma B+\Sigma_{j\sigma}^{\Lambda}$
the full propagator including the self-energy reads
\begin{eqnarray}
{\mathcal{G}}_{\sigma}^{\Lambda}(i\omega)&=&\\&&\!\!\!\!\!\!\!\!\!\!\!\!\!\!
\frac{1}{D_{\sigma}^{\Lambda}(i\omega)}\left( \begin{array}{cc}
\!i\omega\!+\!i\Gamma_R{\rm sgn}(\omega)-\!{V}_{2\sigma}^{\Lambda}&\!\!\!\!\!\!\!\!\!\!\!\!\!\!\!\!-t\\
 -t &\!\!\!\!\!\!\!\!\!\!\!\!\!\!\!\!i\omega\!+\!i\Gamma_L{\rm sgn}(\omega)-\!{V}_{1\sigma}^{\Lambda}\end{array}\! \right)\;,\nonumber
\label{p}
\end{eqnarray}
with the determinant
\begin{eqnarray}
D_{\sigma}^{\Lambda}(i\omega)&=&\\&&\!\!\!\!\!\!\!\!\!\!\!\!\!\!\!\!\!\!\!\!\!\!\![i\omega\!+\!i\Gamma_R{\rm sgn}(\omega)-\!{V}_{2\sigma}^{\Lambda}][i\omega\!+\!i\Gamma_L{\rm sgn}(\omega)-\!{V}_{1\sigma}^{\Lambda}]-t^2\;.\nonumber\phantom{\frac{1}{1}}
\end{eqnarray}
The zeros of $D_{\sigma}^{\Lambda=0}(0)$ for $\Gamma_{L/R}=0$ 
determine the zero-energy levels. For degenerate levels these are responsible for the development of 
the Kondo ridges in the conductance.

Inserting the above expression in Eq.~(\ref{DGLsigma}), the flow equations for the local potential are
\begin{widetext}
\begin{eqnarray}
\frac{d}{d\Lambda}{V}_{1\sigma}^{\Lambda}&=&\frac{U}{\pi}\frac{({V}^{\Lambda}_{1\bar{\sigma}}{V}^{\Lambda}_{2\bar{\sigma}}-t^2){V}^{\Lambda}_{2\bar{\sigma}}+(\Lambda+\Gamma_R)^2{V}^{\Lambda}_{1\bar{\sigma}}}{{[{V}^{\Lambda}_{1\bar{\sigma}}{V}^{\Lambda}_{2\bar{\sigma}}-(\Lambda+\Gamma_L)(\Lambda+\Gamma_R)-t^2]^2+[(\Lambda+\Gamma_R){V}^{\Lambda}_{1\bar{\sigma}}+(\Lambda+\Gamma_L){V}^{\Lambda}_{2\bar{\sigma}}]^2}}\label{locpot}
\end{eqnarray}
\end{widetext}
with an analog equation for ${V}_{2\sigma}^{\Lambda}$ with $(1\leftrightarrow2)$ and 
$(L\leftrightarrow R)$.

We first analyze the symmetric case for $\delta=0$ ($\epsilon_1=\epsilon_2$) and $\Gamma_L=\Gamma_R=\Gamma$. The condition for zero-energy levels 
$V_{\sigma}^2-t^2=0$ implies degenerate solutions for
either $B=0$ and $V=\pm t$ or $B_{\rm ren}=\pm t$ and $V=0$, where the renormalized level position or
Zeeman field replace the bare ones with respect to the noninteracting case.
The flow equation for the local potential (\ref{locpot}) reduces to 
\begin{eqnarray}
\frac{d}{d\Lambda}{V}_{\sigma}^{\Lambda}&=&\frac{U{V}^{\Lambda}_{\bar{\sigma}}}{\pi}\frac{({V}^{\Lambda}_{\bar{\sigma}})^2-t^2+(\Lambda+\Gamma)^2}{[({V}^{\Lambda}_{\bar{\sigma}})^2-t^2-(\Lambda+\Gamma)^2]^2+4({V}^{\Lambda}_{\bar{\sigma}})^2(\Lambda+\Gamma)^2}\nonumber\\
\label{eq:flow}
 &=& \frac{UV_{\bar\sigma}^\Lambda}{\pi}
{\rm Re}\,\frac{1}
  {(\Lambda+\Gamma+it)^2+(V_{\bar\sigma}^\Lambda)^2}
\end{eqnarray}
resembling Eq.~(\ref{eq:diffkondo}) for the single-level quantum dot, except for the presence of 
the finite inter-dot hopping $t$ and a factor $1/2$ in the definition of $\Gamma$.
At $B=0$ the flow equation for $\tilde{V}^{\Lambda}={V^{\Lambda}}- t\, {\rm sgn} (V^{\Lambda})$ 
is characterized by an exponential pinning to zero for 
$|\tilde{V}|\lesssim\, 0.77 U$, inducing a splitting of the central plateau for $\tilde{V}$
in two plateaus of half width for ${V}$ ranging from $V_g=\pm t$ to $\pm (t+0.77 \,U)$.
These are shifted to larger values in Fig.~\ref{fig:fig3} as a consequence of the renormalization of $t$.
The same substitution around the $B=0$ Kondo ridges $V_\sigma^{\Lambda}= t \, {\rm sgn} (V^{\Lambda})+\sigma B^{\Lambda}$ leads to an exponential suppression of the renormalized field as described by 
Eq.~(\ref{bren}) for the single-level dot.
A similar behavior in the $V_g$-$B$ plane is found for the finite-$B$ Kondo ridges. Here the renormalization of $B$ leads to a shift of the position. 
Due to the enhancement of the renormalized magnetic field, the finite-$B$ Kondo ridge
develops at a reduced field for which $B_{\rm ren}=\pm t$. This effect is superposed by the
renormalization of $t$ in Fig.~\ref{fig:fig3} and hardly visible.

We now consider the general asymmetric situation. With
${V}_{1/2\sigma}^{\Lambda}={\bar V}_{\sigma}^{\Lambda}\pm\delta$ and introducing
the asymmetry parameter $\beta$ for the coupling to the leads $\Gamma_{L/R}=\Gamma\pm\beta$,
the flow equation for $\bar{V}_{\sigma}^{\Lambda}$ is 
\begin{widetext}
\begin{eqnarray}
\frac{d}{d\Lambda}\bar{V}_{\sigma}^{\Lambda}&=&\frac{U}{\pi}\frac{
\bar{V}^{\Lambda}_{\bar{\sigma}}[(\bar{V}^{\Lambda}_{\bar{\sigma}})^2-\delta^2-t^2+\beta^2]+(\Lambda+\Gamma)[(\Lambda+\Gamma)\bar{V}^{\Lambda}_{\bar{\sigma}}-2\beta\delta]}
{[(\bar{V}^{\Lambda}_{\bar{\sigma}})^2-\delta^2-t^2+\beta^2-(\Lambda+\Gamma)^2]^2+4[(\Lambda+\Gamma)\bar{V}^{\Lambda}_{\bar{\sigma}}-\beta\delta]^2}\label{b}\;.
\end{eqnarray}
\end{widetext}
Analogously, a flow equation for the renormalization of $\delta$ can be derived, which will not be considered
in the following as it turns out not to significantly affect the results.
For $\delta\neq 0$ and symmetric couplings to the reservoirs $\beta=0$ 
($\Gamma_L=\Gamma_R$), the above equation reduces to 
(\ref{eq:flow}). Kondo plateaus of width $0.77\, U$ are hence obtained in correspondence of degenerate
energy levels at
$\bar{V}=\pm t_{\rm eff}$ for $B=0$ and at $V_g=0$ for $B_{\rm ren}=\pm t_{\rm eff}$.
For asymmetric couplings to the leads ($\beta\neq0$), the above Eq. (\ref{b}) in proximity of the degenerate energy levels can be simplified to
\begin{eqnarray}
\frac{d}{d\Lambda}\bar{V}^\Lambda_{\sigma}&=&\frac{U}{\pi}\frac{\bar{V}^\Lambda_{\bar{\sigma}}-2\beta\delta/\Gamma}{(\Lambda+\Gamma)^2+4(\bar{V}^{\Lambda}_{\bar{\sigma}}-\beta\delta/\Gamma)^2}\;
\end{eqnarray}
in the limit of small asymmetric couplings $\beta\ll\Gamma$.
The terms proportional to $\beta\delta/\Gamma$ induce a \textit{shift} in the effective level position responsible for the $V_g\to -V_g$ symmetry breaking.
For $V_g\gtrsim 0$ an effective magnetic field of strength $B_c+2\beta\delta/\Gamma$ is required to compensate for the shift (corresponding to a larger effective field for $\beta\delta>0$ in Fig.~\ref{fig:fig4}) .
This implies that the finite-$B$ Kondo plateaus in the $V_g$-$B$ plane are not parallel to the $B=0$ axis for 
$\beta\delta\neq0$, see Fig.~\ref{fig:fig4}.

\section{Conclusion}\label{concl}

We studied the linear conductance of a serial quantum dot as a minimal model to describe the influence of the SOI on the spin Kondo effect in presence of a Zeeman field.\cite{galpin}
Without SOI, the linear conductance as a function of an applied gate voltage 
exhibits characteristic Kondo plateaus at finite magnetic field $B$, in addition to the 
ones at $B=0$. Interestingly the finite-$B$ Kondo ridges are bent with respect to the 
$V_g$-axis if the left-right symmetry and the symmetry between the bonding and 
anti-bonding states are broken simultaneously.  
This finding is of importance for the understanding of measurements of
the linear conductance of multi-level quantum dots as a function of
$V_G$ and $B$.
In the presence of SOI the finite-$B$ Kondo ridges disappear; in contrast to the ridges 
at $B=0$, they are not protected by time-reversal symmetry. 

\section*{Acknowledgments}

We are grateful to K. Grove-Rasmussen for inspiration and numerous discussions, and to
C. Karrasch for a critical reading of the manuscript.
This work was supported by the Deutsche Forschungsgemeinschaft 
(FOR 912).


\vfill\eject


\begin{thebibliography}{99}

\bibitem{Hewson} A.C.~Hewson, \textit{The Kondo Problem to Heavy
    Fermions}, (Cambridge University Press, Cambridge, UK, 1993). 
\bibitem{Glazman} L.I.~Glazman and M.E.~Raikh, JETP Lett.\ {\bf 47}, 452 (1988).
\bibitem{Ng} T.K.~Ng and P.A.~Lee, Phys.\ Rev.\ Lett.\ {\bf 61}, 1768 (1988).
\bibitem{Goldhaber} D.~Goldhaber-Gordon, H.~Shtrikman, D.~Mahalu,
  D.~Abusch-Magder, U.~Meirav, and M.A.~Kastner, Nature {\bf
    391}, 156 (1998).
\bibitem{Cronenwett} S.M. Cronenwett, T.H. Oosterkamp, and L.P. Kouwenhoven, 
Science {\bf 281}, 540 (1998).
\bibitem{Schmid} J. Schmid, J. Weis, K. Eberl, and K. von Klitzing, Physica B {\bf 256-258}, 182 (1998).
\bibitem{Wiel} W.~van der Wiel, S.~De Franceschi, T.~ Fujisawa,
  J.M.~Elzerman, S.~Tarucha, and L.P.~Kouwenhoven, Science {\bf
    289}, 2105 (2000).
\bibitem{Meir} Y. Meir and N.S. Wingreen, Phys.\ Rev.\ B {\bf 50}, 4947 (1994). 
\bibitem{thesisB}
J.E. Birkholz, PhD thesis, Universit\"at G\"ottingen (2008).
\bibitem{jens} J. Paaske, A. Andersen, and K. Flensberg, Phys. Rev. B {\bf 82}, 081309(R) (2010).
\bibitem{Pustilnik} M. Pustilnik, Y. Avishai, and K. Kikoin,
Phys. Rev. Lett. {\bf 84}, 1756 (2000).
\bibitem{2} J. Nyg{\aa}rd, D. Cobden, P. E. Lindelof,
Nature {\bf 408}, 342 (2000).
\bibitem{Izumida} W.\ Izumida, O.\ Sakai, and S.\ Tarucha, 
Phys.\ Rev.\ Lett.\ {\bf 87}, 216803 (2001). 
\bibitem{Cox} D.L. Cox and A. Zawadowski, Adv. Phys. {\bf 47}, 599 (1998).
\bibitem{Borda} L. Borda, G. Zarand, W. Hofstetter, B.I. Halperin, and J. von Delft, Phys.
Rev. Lett. {\bf 90}, 026602 (2003).
\bibitem{Winkler} R.~Winkler, \textit{Spin-Orbit Coupling Effects in Two-Dimensional Electron 
and Hole Systems}, Springer, Berlin (2003). 
\bibitem{Mireles} F. Mireles and G. Kirczenow, Phys. Rev. B {\bf 64}, 024426 (2001).
\bibitem{BM}
J.E. Birkholz and V. Meden, J. Phys.: Condensed Matter \textbf{20}, 085226 (2008);
Phys. Rev. B \textbf{79}, 085420 (2009).
\bibitem{KEM}
C. Karrasch, T. Enss, and V. Meden, Phys. Rev. B \textbf{73}, 235337 (2006).
\bibitem{SalmhoferHonerkamp} M.~Salmhofer and C.~Honerkamp, 
Prog.\ Theor.\ Phys.~{\bf 105}, 1 (2001).
\bibitem{AEM}
S. Andergassen, T. Enss, and V. Meden, Phys. Rev. B \textbf{73}, 153308  (2006).
\bibitem{karrasch} 
C. Karrasch, R. Hedden, R. Peters, Th. Pruschke, K. Sch\"onhammer, and V. Meden, 
J. Phys.: Condens. Matter \textbf{20}, 345205 (2008);
C. Karrasch, V. Meden, and K. Sch\"onhammer, Phys. Rev. B \textbf{82}, 125114 (2010).
\bibitem{Severin} S.G. Jakobs, M. Pletyukhov, and H. Schoeller, Phys. Rev. B {\bf 81}, 195109 
(2010).
\bibitem{Oguri} A.~Oguri, J.~Phys.~Soc.~Japan {\bf 70}, 2666 (2001).
\bibitem{Holm} J.V. Holm, H. I. J\o rgensen, K. Grove-Rasmussen, J. Paaske, K. Flensberg, and
P.E. Lindelof, Phys. Rev. B {\bf 77}, 161406(R), 2008.
\bibitem{Hauptmann} J.R. Hauptmann, J. Paaske, and P.E. Lindelof, Nature Physics {\bf 4}, 373 (2008).
\bibitem{kasper}
K. Grove-Rasmussen, private communication.
\bibitem{galpin}
A related model specifically designed for the
description of carbon nanotube multi-level quantum dots was
studied in M.R. Galpin, F.W. Jayatilaka, D.E. Logan, and F.B. Anders, Phys. Rev. {\bf 81}, 075437 (2010).





\end{thebibliography}
\end{document}